\documentclass[fleqn,10pt]{wlscirep}
\title{Electronic thermal conductivity in 2D topological insulator in a HgTe quantum well}

\author[1,*]{G. M. Gusev}
\author[2,3]{ Z. D. Kvon}
\author[1]{A. D. Levin}
\author[2]{E. B. Olshanetsky}
\author[4]{O. E. Raichev}
\author[2]{N. N. Mikhailov}
\author[2]{S. A. Dvoretsky}
\affil[1]{Instituto de F\'{\i}sica da Universidade de S\~ao
Paulo, 135960-170, S\~ao Paulo, SP, Brazil}
\affil[2]{Institute of Semiconductor Physics, Novosibirsk
630090, Russia}
\affil[3]{Novosibirsk State University, Novosibirsk 630090,
Russia}
\affil[4]{Institute of Semiconductor Physics, NAS of
Ukraine, Prospekt Nauki 41, 03028 Kyiv, Ukraine}

\affil[*]{gusev@if.usp.br}

\begin{abstract}
We have measured the differential resistance in a two-dimensional
topological insulator (2DTI) in a HgTe quantum well, as a function
of the applied dc current. The transport near the charge neutrality
point is characterized by a pair of counter propagating gapless
edge modes. In the presence of an electric field, the energy is
transported by counter propagating channels in the opposite direction.
We test a hot carrier effect model and demonstrate that the energy
transfer complies with the Wiedemann Franz law near the charge neutrality point  in the
edge transport regime.
\end{abstract}

\begin{document}

\flushbottom
\maketitle
%
%
\thispagestyle{empty}

\section*{Introduction}

Two-dimensional topological insulators  (2DTI) have attracted considerable attention in condensed matter physics for several
reasons. Among them is the prediction of the existence of almost perfect edge channels in which electrons with opposite spin
move in opposite directions without external magnetic field [1-5].
This hypothesis has been confirmed in micrometer-sized HgTe
quantum wells from observation of the quantized local and nonlocal
resistances [6-8]. Furthermore, the signature of the ballistic
quantum transport is robust in InAs/GaSb quantum wells [9-11],
which is expected to be another variety of a two-dimensional topological
insulator [12].
Another interesting prediction about topological insulators is the
protection of the helical edge states against elastic
backscattering by time reversal symmetry [1-5]. However, this
remarkable property has not yet been experimentally confirmed
[9-13] and identifying the basic mechanism responsible for the
backscattering at the edges of the topological insulators remains
an important unresolved issue.

 Application of novel experimental methods for the investigation
 of the transport properties of 2D TI is of particular interest.
\begin{figure}[ht]
\includegraphics[width=\linewidth]{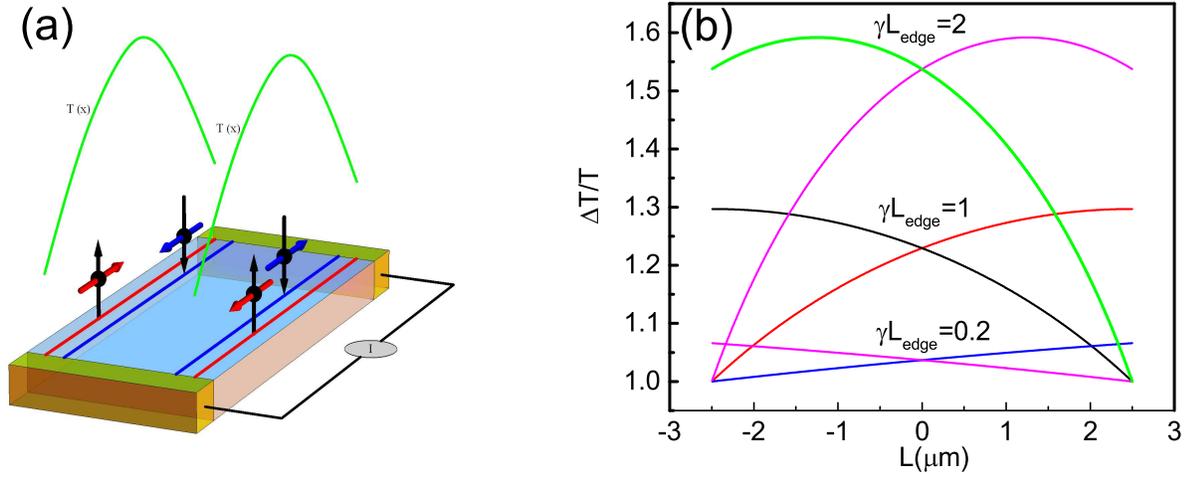}
\caption{(Color online) Schematic drawing of the slab shape sample
with counter propagating spin polarized edge states and the
electron temperature profile near the edge in the diffusive
regime. The temperature profile of the helical states is
calculated from Equation 1 for quasiballistic transport and
for different parameters of $\gamma$ (see text for
explanation).} \label{fig:1}
\end{figure}

Thermal conductivity measurements in metals and semiconductors
have been often used as a powerful tool for probing transport
mechanisms. Within the Landau Fermi liquid theory, electrons carry
both charge and heat, and the relationship between thermal
$\kappa$ and electrical $\sigma$ conductivities is known as the
Wiedemann-Franz (WF) law, $\cal{L}$=$\kappa/\sigma T$. The Lorentz
number takes the universal value
$\cal{L}_{0}$=$(\pi^{2}/3)(k_{B}/e)^2$, where $k_{B}$, $e$ are the
Boltzmann's constant and the electron charge, robust at low
temperature and valid even in the presence of arbitrary
disorder, when the Sommerfeld expansion can be applied. Deviation
from the Wiedemann-Franz law at low temperature in the one dimensional (1D)
case is associated with failure of the Fermi liquid model
[15-18]. In particular, Luttinger liquid effects can be
involved in the breakdown of the Wiedemann-Franz law [19,20].
However, at high temperatures, specific features of the Luttinger
liquid effect begin to broaden and smear out, and a weakly correlated 1D
electron system can be viewed, in a first approximation, as a Fermi
liquid. The Wiedemann-Franz law has been verified in ballistic point contact [21],
and  the experimental results demonstrate a satisfactory agreement
both with the Kelvin-Onsager and Wiedmann-Franz relations. Helical
edge states in topological insulators, however, have properties
that are radically different from those expected in conventional
1D systems. For example, they have a linear dispersion, and,
therefore, the inelastic electron-electron scattering within a
single edge channel is much stronger than the transitions between
the counter propagating edge states (which are rare because of the
topological protection). This leads to counter propagating
heat currents and different carrier thermalization and temperature
profiles [22].

In the present article, we report an experimental study of the
electronic thermal conductivity in band-inverted HgTe-based
quantum wells. At the charge neutrality point, electron transport is dominated by the
edge state currents because the local and nonlocal resistances in
our samples are comparable [13]. We measure non-linear
voltage-current characteristics at different temperatures, and
extract the electron temperature generated by Joule heating using
the 2D TI temperature dependent resistance as its own thermometer.
Assuming that the dissipation of Joule energy in the short
samples occurs through the contacts rather than via phonon emission,
the resulting electronic temperature distribution is universal and
determined by the heating voltage and sample geometry [23]. Figure
\ref{fig:1} shows the temperature profile due to Joule
heating for one channel in accordance with the WF law for
diffusive transport. One can see an inhomogeneous, parabolic profile
of the electronic temperature along the boundary. An unusual
temperature distribution appears in the two-dimensional topological insulators,
where counter propagating electro-thermal flow is expected near
the boundary. We show that overheating of the charge carriers
leads to a nonmonotonic dependence of the differential resistance
on the source-drain bias. We calculate the temperature profile in
our samples and demonstrate that the results are correlated with
the models proposed in [22], describing different temperature
profiles for the counterpropagating edge modes. The
Wiedemann-Franz law is found to be valid near the charge neutrality
point supporting the overheating of the edge state carriers model.
Our findings pave the way for further exploration of quantized thermal
transport in two-dimensional topological insulators.

\section*{Results}

\begin{figure}[ht!]
\includegraphics[width=\linewidth]{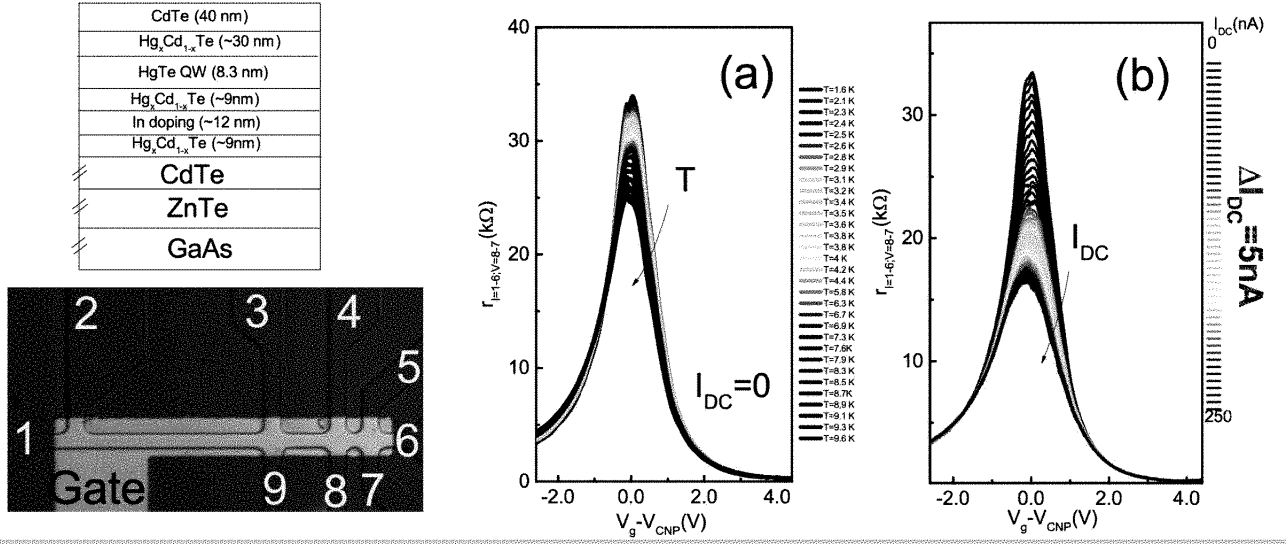}
\caption{(Color online) (a) The differential resistance as a function of
gate voltage and bath temperature. (b) The differential resistance as a
function of gate voltage and the DC current. Left bottom-
schematic structure of the sample. Right bottom - top view of the
sample.} \label{fig:2}
\end{figure}

The $Cd_{0.65}Hg_{0.35}Te/HgTe/Cd_{0.65}Hg_{0.35}Te$ quantum wells
with (013) surface orientations and width $d$ of 8-8.3 nm were
fabricated by molecular beam epitaxy (Figure \ref{fig:2}, left
bottom panel). A detailed description of the sample structure has
been given in Ref 13. The device is designed for multiterminal
measurements and consists of three $3.2 \mu m$ wide consecutive
segments of different length ($3, 9 , 35 \mu m$), and 7 voltage
probes (Figure \ref{fig:2}, right bottom  panel). The ohmic
contacts to the two-dimensional gas were formed by the in-burning
of indium. To fabricate the gate, a dielectric layer containing
100 nm $SiO_{2}$ and 200 nm $Si_{3}Ni_{4}$ was first grown on the
structure using the plasmochemical method followed by a TiAu gate
with the dimensions $62\times8 \mu m^{2}$. The density variation
with gate voltage is $(1.09\pm 0.01)\times 10^{15}
m^{-2}V^{-1}$. Three different devices were studied.

Below we show the results obtained in one representative sample
demonstrating quasi-ballistic behaviour. The differential resistance is studied
when a current flows between contacts 1-6, and voltage is
measured between probes 8 and 7, $r_{1-6,8-7}=dV_{8,7}/dI_{1,6}$. We employ lock-in technique at low-frequency $(\sim 4 Hz)$  using
sufficiently small AC current to prevent self-heating (typically
less than 1 nA). The differential resistance is measured as a
function of the lattice temperature $T_{L}$ and the DC current
$I_{DC}$ applied between the source and drain contacts.
\begin{table}[ht]
\centering
\begin{tabular}{|l|l|l|l|}
\hline

&sample & $\gamma L_{edge}$ & $\cal{L}/\cal{L_{0}}$ \\
\hline
&1&    2 & $1.1\pm0.3$ \\
\hline
&2&   1.5 & $1\pm0.3$ \\
\hline
&3&   1.1 & $1\pm0.2$ \\
\hline
\end{tabular}
\caption{\label{tab1} Parameters of the electron system in HgTe samples at CNP, T=4.2K. Parameters are defined in the text. }
\end{table}

\begin{figure}[ht]
\includegraphics[width=\linewidth]{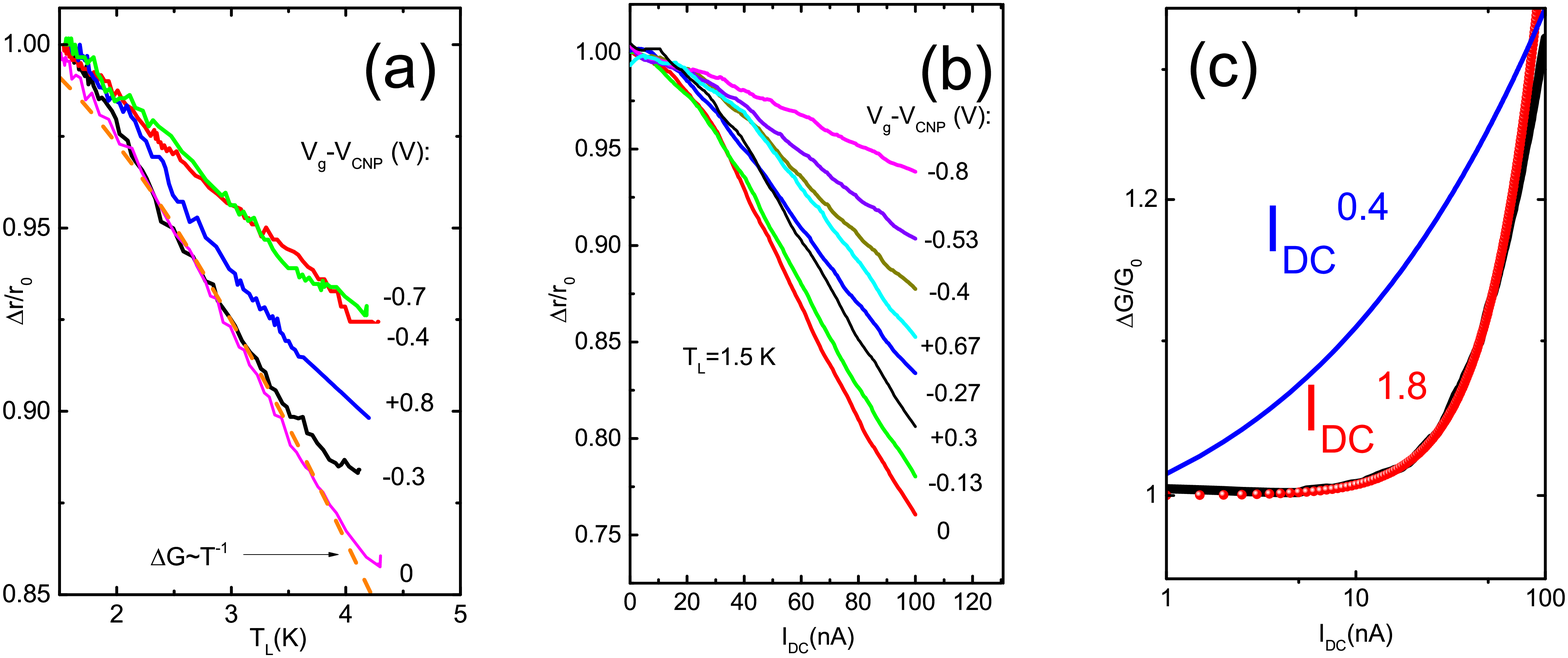}
\caption{(Color online)  (a) Differential resistance as a function of lattice temperature
for different gate voltages. Dashed line -linear T-dependence for
the conductance. (b) Differential resistance as a
function of DC current for different gate voltages. (c) Relative differential conductance as a
function of $I_{DC}$, $T_{L}=1.5K$} \label{fig:3}
\end{figure}

The variation of differential resistance with gate voltage and
lattice (bath) temperature is shown in Figure \ref{fig:2}a. Note
that the minimum probe distance criteria are used to reduce the
Joule heat dissipation via phonon emission. The resistance reveals
a broad peak that is larger than the quantized value $h/2e^{2}$
expected for the ballistic case. For comparison, in Figure \ref{fig:2}b,
we show the variation of differential resistance with gate
voltage and DC source-drain current. One observes a high
similarity between the two plots: the zero bias resistance and the
differential resistance strongly depending on temperature and
$I_{DC}$ respectively. Figures \ref{fig:2}a and b contain
individual curves of $r_{I_{DC}=0}(T)=r_{0}(T)$ and $r(I_{DC})$, allowing for a
detailed comparison between them. Zero bias differential resistance
decreases almost linearly  with temperature $r_{0}(T)\sim
\gamma(V_{g})\times T$, with the temperature coefficient
$\gamma(V_{g})$ strongly depending on gate voltage. Differential resistance depends in a similar way on the bias
current. It decreases with increasing $I_{DC}$, with a nonlinear
coefficient depending on $V_{g}$. The similarity between
$r_{0}(T)$ and $r(I_{DC})$ suggests that the nonmonotonic
differential resistance results from a hot-carrier effect. It is
instructive to linger on the possible mechanisms of the charge
transport along the edge of the two-dimensional topological insulator.
Observation of resistance higher than the quantized value in
relatively long HgTe quantum well samples [6,8,13] supports the
notion that the topological protection in real samples is fragile
due to disorder. There is an increasing number of models that are supposed
to account for the deviation of the experimental resistance values
from the expected quantized value [24].

A promising model has been proposed quite recently [25] claiming
to account for the breakdown of the topological protection. It is
argued that a realistic smooth edge potential results in a
spontaneous breaking of the time-reversal symmetry responsible for
the protection against backscattering. The edge reconstruction
leads to a finite elastic scattering length, and conductance is
described by the equation $G=(e^{2}/h)/(1+\gamma L_{edge})$, where
$\gamma$ is the inverse mean free path length for the edge to edge
scattering, and $L_{edge}$ is the effective length of the edge
channels.

Notice, however, that nonlinear transport is also one of the key
features of a clean Luttinger liquid, due to the so called
zero-bias anomalies resulting from the tunneling of electrons
from the bulk to the wire. It is expected that differential
resistance follows a power law $r\sim V^{\alpha}\sim I ^{\alpha}$,
where $\alpha$ depends on the inter-electron interaction strength
and the number of channels [15-18]. Figure \ref{fig:3}c
displays the relative differential conductivity as a function of
$I_{DC}$. One may see that $r(I_{DC})$ shows a parabolic
dependence rather than the bias dependence of $r\sim I_{DC}^{0.4}$
expected for a single channel. Note, however, that the observation
of the zero-bias anomaly requires very low temperatures because
it scales as a function of $eV_{DC}/kT$ and, at high temperature,
this term becomes smaller than the hot carrier effect.

\begin{figure}[ht]
\includegraphics[width=\linewidth]{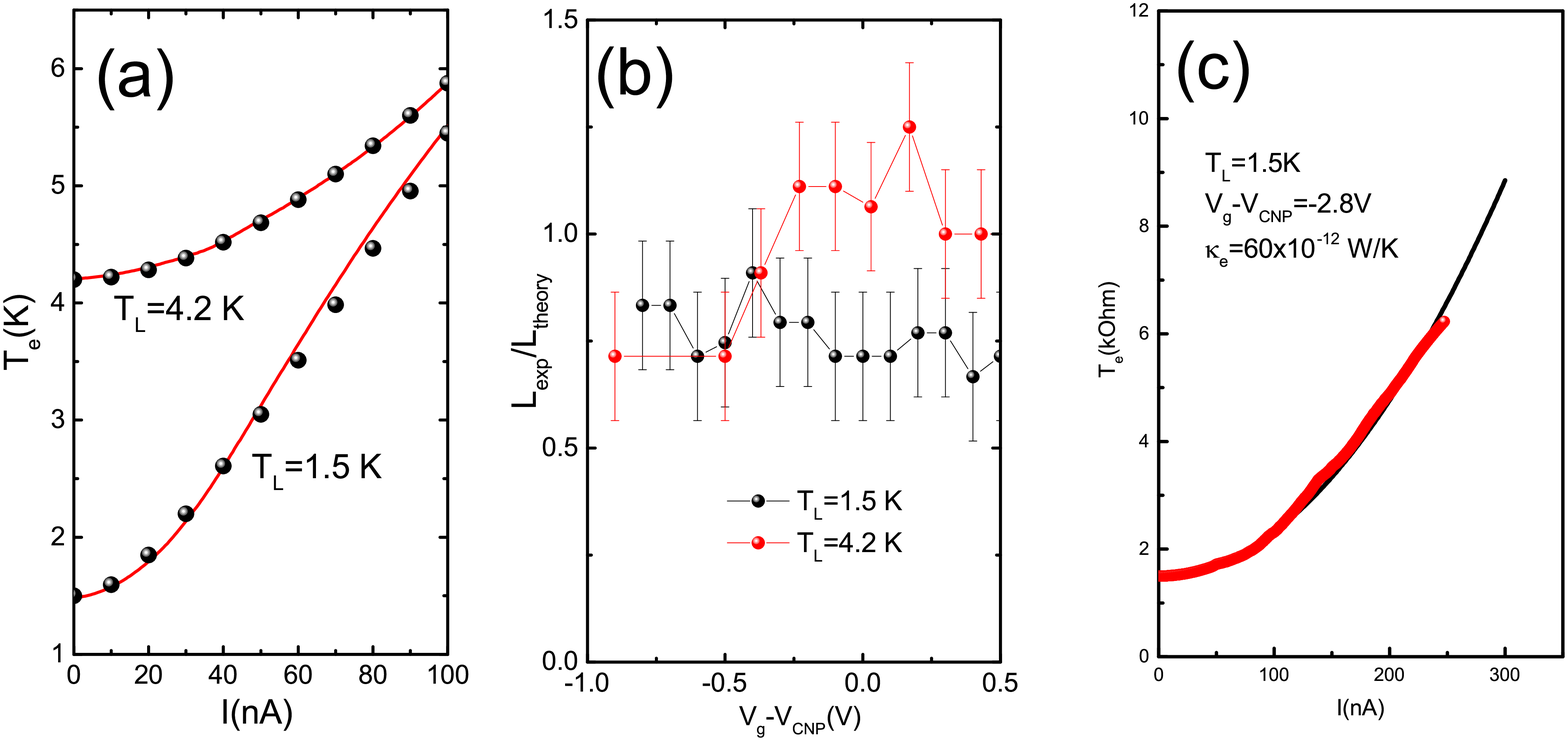}
\caption{(Color online) (a) Dependence of electron temperature,
at the charge neutrality point, on the bias current for two lattice temperatures
$T_{L}$=4.2 and 1.5 K. Points-the averaged temperature found from
equations (1) and (2). (b) Dependence of the ratio of experimental
electron temperature to the averaged temperature found from
Equations (1) and (2) on the gate voltage for two lattice
temperatures. (c) Dependence of electron temperature on the
bias current in the bulk transport regime, $\cal{L}=\cal{L}_{0}$.
} \label{fig:4}
\end{figure}

Before extracting the electron temperature, it is important to
identify the mechanism that is responsible for the temperature
dependence of zero-bias resistance. Firstly, one may suggest
that it is a weak localization effect in one dimensional wires
[26]. Note, however, that one-dimensional Wiedemann-Franz law theories are valid when the
sample resistance is much larger than $h/e^2$ (this is not the
case for our sample when the shortest $L_{edge}$ distance is
used), and they were initially intended for conventional quantum
wires (not for the helical edge states, where the conventional
weak localization mechanisms might not work). To our knowledge,
there is no theory on the Wiedemann-Franz law in helical edge states. Furthermore,
we tried to check if the law $G=G_0-\Delta G$, where $\Delta G \sim
T^{-1}$ is the conductance of the short channel, can account
for the $15\% $ linear resistance decrease
observed in the interval of temperatures from 1.5 K to 4.5 K (Figure
\ref{fig:3}a). Comparing this law with the experimental results, we
found an excellent agreement.

It is worth noting that we also measured differential resistance at higher temperatures.
We  found  that  the  profile $r_{0}(T)$  above $T > 15 K $
fits  very  well  with the  activation law $r_{0} \sim exp(\Delta/2 kT)$, where $\Delta$ is the activation gap. The
 thermally  activated  behavior  of  resistance  above  15  K corresponds  to  a  gap  of  10  meV
  between  the  conduction and  valence  bands  in  the  HgTe  well. The  mobility  gap  can  be
  smaller than  the  energy  gap  due  to  disorder.

Let us now address the transport properties of the helical edge
states. The phase diagram for the conductivity of the helical
liquid in the presence of the disorder and interaction has been
considered theoretically in [17]. Three dominant scattering
mechanisms have been taken into account. The first mechanism, the
inelastic single particle scattering mechanism, denominated as a 1P
process, leads to a change in the chirality of a single incoming
particle. The second important mechanism describes the inelastic
backscattering of two electrons (2P process). Finally, the
authors argue that the backscattering may occur due to electron-electron interactions only [27]. The main results are
summarized as follows:

 \begin{equation}
\Delta G \sim \frac{e^{2}}{h}L_{edge} T^{-2K-2},  K > 2/3
\end{equation}
 \begin{equation}
\Delta G \sim \frac{e^{2}}{h}L_{edge} T^{-8K+2},  K < 2/3
\end{equation}

where $K$ is the Luttinger liquid parameter. It is important to
note that, at $K > 2/3$, transport properties are dominated by 1P
scattering, and below $K = 2/3$, the 2P process becomes important.
In addition it is expected that, below $K < 3/8$, the
localization of the helical edge states takes place. Assuming the
Luttinger parameter $K\approx3/8$ (2P process), which corresponds to the weak
coupling regime, we obtain a good agreement with the experimental
dependence $\Delta G \sim \frac{e^{2}}{h}L T^{-1}$. Intriguingly,
this parameter value also marks the transition from the localized
to the delocalized regime in the transport of the disordered
helical liquid. However, the study of the localization effects is
out of scope of the present paper. It is worth noting that a
previous study of strongly disordered topological insulators
revealed a practically temperature independent resistance [13]
with, however, some tendency to localization behaviour.

It is also worth noting that the electron-phonon scattering
predicted a $T^3$ dependence of the scattering rate [18], which disagrees with our observations.

As our sample resistance strongly depends on the electron
temperature, this property can be used as a thermometer. Moreover,
since the bulk conductivity is much smaller than the edge
contribution (which is testified by a pronounced nonlocal effect
[13]), the thermoconductivity must be dominated by the edge state
transport and model [22] is valid.

\section*{Discussion}
A quantitative check of the heat distribution in the sample
provides further support for the hypothesis that the nonlinear
effect results from Joule heating and electron thermal
conductivity. A weak electron-phonon scattering is allowed in the
presence of Rashba spin orbit coupling. However, because the sample is short,
phonon emission does not occur at this distances (for details see Supplementary material)
In addition, our samples are not
ballistic as their peak resistance is larger than the
quantized value (Figure \ref{fig:2}). Independently of a
particular mechanism responsible for the backscattering (except for
the electron-phonon coupling), in our samples, we obtain
$L_{edge}\gamma>> 1$ and we argue that the local temperature
profile is controlled by the balance between Joule heating and
electron thermal conductance. We argue that Joule heat is
transferred by electrons and then dissipated via the electron
diffusion into the bulk contacts outside the gate region. On the
other hand, the inelastic electron-electron scattering within a
single edge channel is free from these restrictions and,
therefore, is much stronger. Moreover, such kind of scattering
sweeps a large phase space, especially if the edge state velocity
is energy-independent so that the momentum and the energy
conservation laws are satisfied simultaneously for any two
electrons participating in the collisions. Therefore, the
temperature profile can be calculated, taking into account the
boundary conditions $T(0)=T(L)=T_{0}$ and Joule heat due to
the current flow [22]:

\begin{equation}
\frac{T_{\pm}(x)}{T_{0}}=\sqrt{1+\frac{3}{\pi^{2}}\left[1+\frac{1\mp 2x/L_{edge}}{2/\gamma L_{edge}}\right]\frac{1\pm 2x/L_{edge}}{2/\gamma L_{edge}}\left(\frac{I_{DC}}{I_{0}}\right)^{2}}
\end{equation}

where $I_{0}=ek_{B}T_{0}/\pi \hbar$  is the current associated
with lattice temperature. Note that this equation has been
derived within the approximation that all the heat is transferred
by the electrons. In this case, the carrier thermal conductivity
for each branch is given by $\kappa=\gamma G_{therm}/2$, where
$G_{therm}=\frac{\pi k_{B}^{2}T}{6\hbar}$ is the quantized thermal
conductance, and $k_{B}$ is the Boltzmann constant [22]. Indeed, it
was supposed that the ratio between the thermal and electrical
conductivity for each branch obeys the Wiedemann Franz law
$\frac{\kappa}{\sigma T}=\cal{L}_{0}$. The ratio
$\frac{I_{DC}}{I_{0}}\sim \frac{G}{G_{therm}}$ can be used as an
adjustable parameter in order to check the Lorentz ratio. The
temperature profile strongly depends on parameter $\gamma$.
Figure \ref{fig:1}b shows the temperature difference along the
sample for three different regimes: ballistic
$L_{edge}\gamma=0.2$, quassiballistic $L_{edge}\gamma =1$, and
intermediate between ballistic and diffusive $L_{edge}\gamma =2$
regimes and the same ratio $\frac{I_{DC}}{I_{0}}=1.5$. One can see that,
due to the unidirectional character of thermal flow, for each
mode, the temperature in each branch increases with distance.
The heat is then dissipated in the contact region with the
two-dimensional electron gas and, subsequently, transferred to the
lattice. Therefore, the electron temperature abruptly decreases
when the edge channel strikes the contact region outside the gate.
In the ballistic regime, the temperature profile increases linearly
with distance, while in the diffusive regime, the temperature
profile has a parabolic shape [22]. In our samples $L_{edge}\gamma
=2$, therefore, we expect a parabolic temperature profile.

 From the temperature profile,
we compute the average electron temperature as a function of the
applied bias current:

\begin{equation}
T_{e}(I_{DC})=\frac{1}{L_{edge}}\int \frac{(T_{+}(x)+T_{-}(x))}{2}dx
\end{equation}

The low bias resistance versus lattice temperature dependence is
thus used to determine the electron temperature $T_{e}$ from the
measured differential resistance $r(I_{DC})$ versus applied bias
current. However, we cannot discriminate between the temperatures
$T_{+}$ and $T_{-}$ as we cannot discriminate between the
contribution to the conductivity from each helical edge state.

The dependence of $T_{e}(V_{DC})$  is compared with  our
experimental data using Equations (3) and (4). We perform
curve fitting and extract the ratio of the experimental
temperature to the average temperature found from Equations (3)
and (4) as a function of the gate voltage and the two temperatures.
All approximation considered, the agreement is good. The
dependence of $T_{e}(V_{DC})$  is compared with our experimental
data using the Lorentz number as an adjustable parameter.

Fig \ref{fig:4}b shows the Lorentz ratio as a function of gate
voltage. The ratio $\cal{L}/\cal{L_{0}}$ is close to one at both
temperatures when the Fermi level stays in the insulating gap
and transport is determined by the edge states. Good agreement
with the Wiedemann-Franz law strongly supports the assumption that
the thermal transport in a 2D topological insulator occurs via the
edge states.

It is worth noting that in the bulk transport regime the
temperature profile becomes universal and depends on the
conductivity and the applied current:

\begin{equation}
T_{e}^{2}= T_{0}^{2}+ \frac{3}{4\pi^{2}}\left(\frac{eI_{DC}}{G}\right)^{2}\left(1-\frac{4x^{2}}{L^{2}}\right)
\end{equation}

where $L$ is now the distance between probes. Indeed this universality is provided by the Wiedemann-Franz law
[23]. The average temperature change is given by $\Delta
T=\frac{PL^{2}}{12\kappa_{e}}$, where $P=J^{2}/\sigma LW$ (J is the current density, $\sigma$ is the conductivity) is the
Joule heating and $\kappa_{e}$ is the bulk electron thermal
conductivity.

We also studied the nonlinear differential resistance in the
bulk transport regime $V_{g}-V_{CNP} > \mid2.8 V\mid$. As the
temperature dependence of the resistance is very weak in this
regime, the temperature distribution is described by Equation
5. We recover the Wiedmann-Franz law, which gives support to our
hypothesis that, for short distances, Joule heat is transferred
only by electrons and justifies our analysis (Fig \ref{fig:4}c). Note that,
for the bulk two-dimensional transport, the thermal conductivity is given by
$\kappa_{e}=\frac{W}{12L}G_{th}$, where W is the width of the sample, and $G_{th}$
is the thermal conductance.

In summary, we have measured the nonlinear differential resistance
together with the low bias resistance in a 2D TI in HgTe quantum
wells. We attribute the nonlinear effects to electron heating
in the helical one dimensional edge states.  We demonstrate that
thermal electron conductivity is determined by edge state
transport. By plotting the temperature profile as a function of
the current bias, we extract the electron thermal conductivity near
the CNP. The Wiedmann-Franz law is valid for both the edge states
and the bulk transport regime.

\section*{Methods}

The samples studied here were grown by molecular
beam epitaxy MBE [28]. A schematic section through the structure
is shown in Fig. 2. Unlike the structures investigated
previously [6-7] in which a (100) surface was used for the
MBE of the quantum wells, we have used a (013) surface.
The quantum wells grown on it can have better quality than in the case of (100), and for
that reason, we used a (013) surface in the present
study[13].
The differential resistance was measured by use of an AC technique of
applying a sinusoidal signal superimposed on a DC bias
to the sample. Then a lock-in amplifier was used to obtain the AC voltage across and the
AC current through the device.
The nonlinear differential
resistance was observed to decrease with increasing electrical
bias current, ($dr/dI_{DC} < 0$) , and was also observed to decrease with
increasing lattice temperature
($dr/dT_{L} < 0$). Assuming that the change in the resistance with an
applied electric field can be described in terms of electric field
induced electron heating, temperature T in Eq. 2 in the main text
can be replaced by the electron temperature $T_{e}$.  Therefore,
$T_{e}$ can be determined by comparing the relative resistances measured as functions of the
lattice temperature $T_{L}$ and the applied electric current $I_{DC}$ :
\begin{equation}
\Delta r(T_{L})/r(T_{L_{0}})= \Delta r(I_{DC})/r_{0}
\end{equation}
where $T_{L_{0}}=1.5 K$ or 4.2 K. The electron temperature as a function of DC current was obtained from the curves shown in Figure 3 of the main text.

\section*{Acknowledgements}
The financial support of this work by the Russian Science
Foundation (Grant No.16-12-10041), FAPESP (Brazil),
and CNPq (Brazil) is acknowledged.

\section*{Author contributions statement}

G. M. G., A. D. L., Z. D. K and E. B. O. performed the experiment,
N. N. M. and S. A. D. synthesized the crystals, G. M. G., O.E.R.
and A. D. L. provided the theoretical framework, G. M. G. wrote
the manuscript with inputs from all authors. G. M. G  and Z. D. K.
supervised the work.  All authors reviewed the manuscript.

\section*{Additional information}

Supplementary information accompanies this paper at https://.
To include, in this order: \textbf{Accession codes} (where applicable); \textbf{Competing financial interests} (The authors declare that they have no competing interests).

The corresponding author is responsible for submitting a \href{http://www.nature.com/srep/policies/index.html#competing}{competing financial interests statement} on behalf of all authors of the paper.


\begin{thebibliography}{24}
\bibitem{1}
C. L. Kane and E. J. Mele, Phys. Rev. Lett. {\bf 95}, 146802 (2005).
\bibitem{2}
B. A. Bernevig, T. L. Hughes, and S. C. Zhang, Science {\bf 314},
1757 (2006).
\bibitem{3}
M. Z. Hasan and C. L. Kane, Rev. Mod. Phys. {\bf 82}, 2045 (2010);
\bibitem{4}
X.-L.Qi, S.-C. Zhang, Rev. Mod. Phys. {\bf 83}, 1057 (2011).
\bibitem{5}
Giacomo Dolcetto, Maura Sassetti, Thomas L. Schmidt, Rivista del Nuovo Cimento 39, 113 (2016)
\bibitem{6}
M. K$\ddot{o}$nig, S. Wiedmann, C. Brune, A. Roth, H. Buhmann, L. W. Molenkamp,
X.-L. Qi, and S.-C. Zhang, Science {\bf 318}, 766 (2007).
\bibitem{7}
A. Roth, C. Br\"une, H. Buhmann, L. W. Molenkamp, J. Maciejko,
X.-L. Qi, and S.-C. Zhang, Science {\bf 325}, 294 (2009).

\bibitem{8}
E. B. Olshanetsky, Z. D. Kvon, G. M. Gusev, A. D. Levin, O. E. Raichev, N. N. Mikhailov,
and S. A. Dvoretsky, Phys.Rev.Lett.{\bf 114}, 126802 (2015).

\bibitem{9}
I. Knez, R.-R. Du, and G. Sullivan, Phys. Rev. Lett. {\bf 107}, 136603 (2011); I. Knez,
C. T. Rettner, S.-H. Yang, S. S. P. Parkin, L. Du, R.-R. Du, and G. Sullivan,
Phys. Rev. Lett. {\bf 112}, 026602 (2014).

\bibitem{10}
F. Nichele, A. N. Pal, P. Pietsch, T. Ihn, K. Ensslin, C. Charpentier, and W. Wegscheider,
Phys. Rev. Lett. {\bf 112}, 036802 (2014).

\bibitem{11}
K. Suzuki, Y. Harada, K. Onomitsu, and K. Muraki, Phys. Rev. B {\bf 91}, 245309 (2015).

\bibitem{12}
F. Qu, A. J. Beukman, S. Nadj-Perge, M. Wimmer, B.-
M. Nguyen, W. Yi, J. Thorp, M. Sokolich, A. A. Kiselev,
M. J. Manfra, C. M. Marcus, and L. P. Kouwenhoven,
Phys. Rev. Lett. 115, 036803 (2015).

\bibitem{13}
G. M. Gusev, Z. D. Kvon, O.A.Shegai, N. N. Mikhailov, S. A. Dvoretsky
and J. C. Portal, Phys. Rev. B {\bf 84}, 121202(R), (2011);
G.M.Gusev, Z. D. Kvon, E. B. Olshanetsky, A. D. Levin, Y. Krupko, J. C. Portal, N. N. Mikhailov, and S. A. Dvoretsky,
 Phys. Rev. B,  89, 125305 (2014).

\bibitem{14}
Thomas L. Schmidt, Stephan Rachel, Felix von Oppen, and Leonid I.
Glazman, Phys. Rev. Lett. {\bf 108}, 156402, (2012).

\bibitem{15}
Francois Crepin, Jan Car Budich, Fabrizio Dolcini, Patrik Recher,
and Bjorn Trauzettel1, Phys. Rev. B {\bf 86}, 121106(R), (2012).
\bibitem{16}
C. Wu, B. A. Bernevig, and S. Zhang, Phys. Rev. Lett., {\bf 96},
106401 (2006).
\bibitem{17}
N. Kainaris, I. V. Gornyi, S. T. Carr, and A. D. Mirlin,
 Phys. Rev. B 90, 075118 (2014).
\bibitem{18}
T. L. Schmidt, S. Rachel, F. von Oppen, and L. I. Glazman,
 Phys. Rev. Lett. 108, 156402 (2012).

\bibitem{19}
M.-R. Li and E. Orignac, Europhys. Lett. 60, 432 (2002).
\bibitem{20}
C. L. Kane and M. P. A. Fisher, Phys. Rev. Lett. 76, 3192
(1996); A. Houghton, S. Lee, and B. J. Marston, Phys.
Rev. B 65, 220503 (2002); M. G. Vavilov and A. D. Stone,
Phys. Rev. B 72, 205107 (2005); B. Kubala, J. Ko$\ddot{o}$nig, and
J. Pekola, Phys. Rev. Lett. 100, 066801 (2008).
\bibitem{21}
L. W. Molenkamp, Th. Gravier, H. van Houte, O. J. A. Buijk, M. A. A. Mabesoon, C. T. Foxon,
Phys. Rev. Lett. 68, 3765 (1992).
\bibitem{22}
Marcelo A. Kuroda and Jean-Pierre Leburton,
Phys. Rev. Lett. 101, 256805 (2008).
\bibitem{23}
V.I. Kozub and A.M. Rudin,
Phys. Rev. B 52, 7853 (1995).
\bibitem{24}
J. Maciejko, C. Liu, Y. Oreg, X.-L. Qi, C.Wu, and S.-C. Zhang,
Phys. Rev. Lett. 102, 256803 (2009); Y. Tanaka, A. Furusaki, and K. A. Matveev, Phys. Rev. Lett.
106, 236402 (2011); V. Cheianov and L. I. Glazman, Phys. Rev. Lett. 110, 206803
(2013); A. Strom, H. Johannesson, and G. I. Japaridze, Phys. Rev. Lett.
104, 256804 (2010); N. Lezmy, Y. Oreg, and M. Berkooz, Phys. Rev. B 85, 235304
(2012); F. Crepin, J. C. Budich, F. Dolcini, P. Recher, and B. Trauzettel, Phys. Rev. B 86, 121106(R) (2012);
J. I. Vayrynen, M. Goldstein, and L. I. Glazman, Phys. Rev. Lett.
110, 216402 (2013); Anders Mathias Lunde and Gloria Platero, Phys. Rev. B ,88, 115411 (2013);
Sven Essert, Viktor Krueckl, and Klaus Richter, Phys. Rev. B 92, 205306 (2015).
\bibitem{25}
Jianhui Wang, Yigal Meir, and Yuval Gefen,
Phys. Rev. Lett. 118, 046801 (2017).
\bibitem{26}
E. Akkermans and G. Montambaux, Mesoscopic Physics of Electrons
and Photons (Cambridge University Press, Cambridge,
2007).
\bibitem{27}
S. S. Kubakaddi, Phys. Rev. B 75, 075309 (2007).
\bibitem{28}
N.N. Mikhailov, R.N. Smirnov, S.A. Dvoretsky, Yu.G. Sidorov, V.A. Shvets, E.V. Spesivtsev, S.V. Rykhlitski,
Int. J. Nanotechnology 3, 120 (2006).

\end{thebibliography}
\end{document}